# High temperature deformation mechanisms in monolithic 3YTZP and 3YTZP containing single wall carbon nanotubes

Miguel Castillo-Rodríguez[1*], Antonio Muñoz[2] and Arturo Domínguez-Rodríguez[2**]

[1]Instituto de Ciencia de Materiales de Sevilla, CSIC-Universidad de Sevilla, Avda. Américo Vespucio 49, 41092 Sevilla, Spain

[2]Departamento de Física de la Materia Condensada, Facultad de Física, Universidad de Sevilla, Apartado 1065, 41080 (Sevilla), Spain

**Abstract:** Monolithic 3YTZP and 3YTZP containing 2.5 vol.% of single-walled carbon nanotubes (SWCNT) were fabricated by Spark Plasma Sintering (SPS) at 1250 ºC. Microstructural characterization of the as-fabricated 3YTZP/SWCNTs composite shows a homogeneous CNTs dispersion throughout of ceramic matrix. The specimens have been crept at temperatures between 1100 and 1200 ºC in order to investigate the influence of the SWCNTs addition on high temperature deformation mechanisms in zirconia. Slightly higher stress exponent values are found for 3YTZP/SWCNTs nanocomposites (n~2.5) compared to monolithic 3YTZP (n~2.0). However, the activation energy in 3YTZP (Q = 715 ± 60 kJ/mol) experiences a reduction of about 25% by the addition of 2.5 vol.% of SWCNTs (Q = 540 ± 40 kJ/mol). Scanning electron microscopy studies indicate that there is no microstructural evolution in crept specimens, and Raman spectroscopy measurements show that SWCNTs preserved their integrity during the creep tests. All these results seem to indicate that the high temperature deformation mechanism is grain boundary sliding (GBS) accommodated by grain-boundary diffusion, which is influenced by yttrium segregation and the presence of SWCNTs at the grain boundary.

*Corresponding author: miguelcr@us.es. Instituto de Ciencia de Materiales de Sevilla, CSIC-Universidad de Sevilla, Avda. Américo Vespucio 49, 41092 Sevilla (Spain). Tel: Int-34-95 455 09 64, Fax: Int-34-95 461 20 97.

**Fellow of the ACS

## 1. Introduction

Although ceramics have high strength and hardness together with excellent thermal-chemical stability, their brittleness at low temperatures restricts their use as structural materials.[1] Since superplasticity was discovered in yttria stabilized tetragonal $Zr_2O$ polycrystals (YSZP),[2] many efforts have been devoted to study this system, where grain boundary sliding GBS is recognized as being the primary deformation mechanism responsible for the high temperature superplastic behaviour accommodated by cationic diffusion throughout the lattice.[3-11] Extensive high temperature creep data are available in the literature, where experimental data are analyzed using the standard phenomenological creep equation:

$$\dot{\varepsilon} = A \frac{\sigma^n}{d^p} \exp\left(-\frac{Q}{KT}\right) \qquad (1)$$

being $A$ a stress and temperature independent term reflecting the dependence of the strain rate on the microstructural features of the material (composition, amount and physical properties of glassy phases, grain morphology, etc.), $\sigma$ the applied stress, $d$ the grain size, $k$ the Boltzmann's constant, and T the absolute temperature. The parameters n, p, and Q (generically known as creep parameters) are, respectively, the stress and grain size exponents, and the apparent activation energy for creep. These creep parameters vary significantly (n between 1 and 5, p between 1 and 3 and Q between 450 and 700 kJ/mol) for nominally similar materials tested under identical experimental conditions.[5-6] Despite the high dispersion in results, it is observed a clear dependence on the YSZP residual impurity content. Low purity YSZP (residual impurity content above 0.1 wt%) exhibits constant values of the creep parameters over the whole stress range (n~2, p~2 and Q~ 500 kJ/mol), whereas in purer materials two different regimes is observed. On the one hand, high stress regime with similar n, p and Q values to those found in low purity materials; and on the other hand, low stress regime where creep parameters could increase up to n~5, p~3 and Q~700 kJ/mol.[5-6] The high temperature plasticity of this material was successfully explained in terms of a threshold stress, whose origin is based upon yttrium segregation at the grain boundaries.[7-11]

Superplasticity of ceramic materials is of great interest both from a scientific point of view and also because of potential applications in industry. However, its brittleness at low temperatures imposes limitations on its industrial uses. Hence, reinforcement of multifunctional nano-structured ceramics has attracted the attention of many investigations since last decades, with the goal of enhancing fracture toughness at low temperatures. Fibers, whiskers, and more recently carbon nanotubes (CNTs) have been incorporated to ceramic matrix as reinforcing elements.[12-15] Among them, CNTs possess extraordinary mechanical, electrical and thermal properties.[16-18] Thus, the combination of these nanotubes with a ceramic matrix could potentially create composites with exceptional toughness and creep resistance. Moreover, the development of proper powder processing routines to get better CNTs dispersion throughout a ceramic matrix[19,20] and the use of new sintering techniques like spark plasma sintering (SPS) allow obtaining fully dense composites preserving CNT integrity. Recently, a study on the influence of the processing route on the carbon nanotubes dispersion in a 3YTZP ceramic matrix was reported.[21] This work showed that aqueous colloidal processing (ACP) using an ultrasonic probe and the freeze-drying technique led to an optimal CNTs distribution on the 3YTZP ceramic matrix.

Most studies on the high temperature mechanical properties of 3YTZP/CNTs composites are related to mechanical spectroscopy measurements.[22-24] In all of them the incorporation of MWCNTs in a 3YTZP ceramic matrix generated a decrease in the internal friction which was attributed to the pinning effect of MWCNTs on the grain boundary sliding. These results were associated to a higher creep resistance at high temperatures. However, creep studies on 3YTZP/CNTs composites reported in literature are very scarce. Castillo-Rodríguez et al.[21] reported that the incorporation of SWCNTs in a 3YTZP ceramic matrix has a softening effect; since 3YTZP/SWCNTs composites exhibit a creep resistance about 60 times lower than that corresponding to the monolithic 3YTZP. This indicates that the role of SWCNTs consists on making GBS easier during deformation instead of acting as a reinforcing element. The opposite effect was found by Mazaheri et al.[25]. They performed creep tests on zirconia containing different quantities of MWCNTs instead of SWCNTs, and they reported a creep rate decrease of three orders of magnitude for a composite with 5 wt.% of MWCNTs concentration compared to monolithic zirconia. This discrepancy could be due to the use of different type of CNTs that may influence differently the high temperature mechanical behaviour of 3YTZP ceramic matrix.

The present work is devoted to investigate the high temperature mechanical behaviour of monolithic 3YTZP and 3YTZP containing 2.5 vol.% of SWCNTs. The aim is to study the influence of the SWCNTs addition on high temperature deformation mechanisms in zirconia. Experimental data are showed and discussed. It is worth emphasizing the novelty of such data, since there are no systematic creep studies reported in literature devoted to investigate the high deformation mechanisms in 3YTZP/CNTs composites.

## 2. Experimental Procedure

*2.1. Starting materials, powder processing and sintering*

3 mol% yttria stabilized zirconia 3YTZP powder (Nanostructured and Amorphous Materials Inc., Houston, TX), with 40 nm average particle size and 99% purity, was annealed in air at 1250 ºC for one hour and subsequently ball milled for 3 hours with a frequency of 25 vibrations/s (Model MM200, Retsch GmbH, Haan, Germany).

Commercially available 90% purified SWCNTs, with a typical bundle length of 0.5-1.5 μm and diameter between 4-5 nm, were provided by Carbon Solutions Inc. (Riverside, CA). They were COOH-functionalizated following the same routine than Poyato *et al.*[26], although at the end of the routine the acid-treated SWCNTs were freeze dried instead of being dried on hot plate.

Among all the powder processing routines described in Ref. [21], it has been selected that providing the optimal CNTs dispersion throughout the 3YTZP ceramic matrix, which is briefly described as follows: Aqueous colloidal processing ACP[20] where SWCNTs and 3YTZP powder were sonicated separately in two high pH solutions (distilled water + $NH_3$ solution until a pH=12) for 1 hour in an ultrasonic probe (Model KT-600, Kontes Inc., Vineland, New Jersey). Later, they were suspended in the same high pH solution, and sonicated again for 30 min. After mixing 3YTZP powder and SWCNTs in the high pH solution, they were frozen immediately by immersing the solution into liquid nitrogen in order to avoid the 3YTZP powder decantation. Finally the blend was freeze dried.

Composite powders (3YTZP + 2.5% vol.% of SWCNTs) were sintered by Spark Plasma Sintering (Model Dr Sinter 1050, Sumitomo Coal Mining Co. LTD, Tokyo, Japan) in a graphite die (10 mm inner diameter) at 1250 ºC, with a constant uniaxial pressure of 75 MPa for 5 min and in ~ $10^{-2}$ mbar vacuum. The heating and cooling rates were 300 ºC min$^{-1}$ and 50 ºC min$^{-1}$, respectively. For the sake of comparison, monolithic 3YTZP was sintered using the same conditions. Rather dense specimens (relative density > 99%) were obtained. Densities of the sintered specimens were measured by Archimedes' method taking a SWCNT density of 1.3 g/cm$^3$, as it was given by the providers.

Impurity concentrations in as-sintered monolithic 3YTZP has been determined by PIXE (particle induced X-ray emission) at the Centro Nacional de Aceleradores (CNA, Seville, Spain), and a total concentration of impurities (Ti, Al, Si, Fe, Na and Cl) higher than 0.1 wt.% was detected.

*2.2. Microstructural characterization*

The microstructure of the samples was examined by light microscopy (Model Leica DMRE, Leica Microsystems GmbH, Germany), by conventional scanning electron microscopy SEM (Model JEOL 6460LV, JEOL USA Inc., MA, USA) and by high resolution scanning electron microscopy HRSEM (Model HITACHI S5200, Hitachi High-Technologies Corporation, Tokyo, Japan). Cross-section specimens were prepared, where surfaces were first ground and then polished with diamond paste down to 1 μm. To reveal 3YTZP grain boundaries, polished surfaces were thermally etched at 1100 ºC for 15 minutes in air or in Ar in case of 3YTZP/CNTs composites to avoid the combustion of SWCNTs. The grain morphology characterization was made by measuring the equivalent planar diameter $d=(4\times area/\pi)^{1/2}$, and the shape factor

F=4π×area/(perimeter)² from HRSEM micrographs. These same parameters were evaluated to characterize CNTs agglomerates. Surface density of SWCNTs agglomerates was estimated from the area fraction covered by them in low magnification SEM micrographs and also by means of light microscope observations.

Fracture cross-sections from composite specimens were also investigated by HRSEM, and by Raman spectroscopy to study SWCNTs integrity after sintering and after creep tests, using a dispersive microscope (Model LabRAm Horiba Jobin Yvon, Horiba Ltd, Kyoto, Japan). Measurements were done with a green laser (He–Ne 532.14 nm), 20-mW, 600 g/mm grating and without filter. A 100× objective and a confocal pinhole of 100 μm were used. The Raman spectrometer was calibrated using a silicon wafer.

*2.3. Mechanical tests*

Uniaxial compression creep tests were performed on a prototype creep machine[27] at temperatures between 1100 and 1200 ºC, and stresses ranging between 5 and 230 MPa. They were carried out in a controlled argon atmosphere to avoid the combustion of SWCNTs. Samples were cut and ground as parallelepipeds of approximate dimensions $5 \times 2.5 \times 2.5$ mm³. Once reached the testing temperature and after thermal stabilisation of the system, a constant load were applied to the specimen. After a brief transitory state, a steady state characterized by a constant strain rate $\dot{\varepsilon}$ is observed. Successive stress or temperature changes were performed obtaining new steady states for each experimental condition. The creep curves were analyzed using the standard phenomenological creep equation (Eq. 1). Creep parameters together with the microstructural study after deformation provide information about the mechanisms operating during the deformation and help the proper understanding of the high temperature mechanical behaviour of materials.

## 3. Results and Discussion

*3.1. Microstructural characterization*

Fig. 1a and 1b show the microstructure of the sintered specimens, monolithic 3YTZP and 3YTZP/SWCNTs composite respectively. The former corresponds to a polished and thermally etched surface and the latter to a fracture surface. 3YTZP grains are faceted with sharp triple points and in case of nanocomposites, CNTs are forming bundles surrounding 3YTZP grains. Fig. 1c is a HRSEM micrograph of the 3YTZP/SWCNTs composite after creep test. It is worth emphasizing that CNTs are clearly visible surrounding 3YTZP grains, indicating that they kept their integrity during creep tests. CNTs agglomerates in composites have been characterized, obtaining an agglomerate surface density of 0.3± 0.1 %, an average size of 0.5 ± 0.3 µm and a shape factor of 0.83 ± 0.15. These values are quite similar to those exhibited by composites previously processed by Castillo *et al.*[21] following the same routine (sample 1bFD-probe in Ref [21]).

Table I shows the relative density values higher than 99% measured in all sintered specimens. Morphological parameters (equivalent planar diameter and shape factor), calculated in both monolithic 3YTZP and 3YTZP/SWCNTs composite, are also shown in Table I. The former exhibits 3YTZP grains with an equivalent planar diameter d of about 0.27 µm, whereas a slightly smaller value, d=0.20 µm, is obtained for the composite. This is evidence that SWCNTs surrounding 3YTZP grains can inhibit the ceramic grain growth during sintering. This effect was also observed by other authors.[28,29] Concerning to the shape factor, all the specimens exhibit a value close to 0.7. It is worth emphasizing that these microstructures do not evolve during creep tests since similar morphological parameter values are obtained for specimens before and after creep experiments.

The CNTs integrity of specimens after sintering and testing has been studied by Raman spectroscopy. Fig. 2 shows Raman spectra performed in both as-sintered and deformed 3YTZP/SWCNT composites. Raman spectra measured in monolithic 3YTZP ceramic and in SWCNTs are also included for comparison. Composites show on the one hand peaks at 165, 260, 320, 465, 610 and 643 cm$^{-1}$ which corresponds to the six Raman bands predicted for theoretical tetragonal zirconia. On the other hand, they also show the typical mode bands due to the presence of CNTs: (i) radial breathing mode bands located about 150-200 cm$^{-1}$, and (ii) the G mode corresponding to the tangential shear mode of carbon atoms (~1550-1600 cm$^{-1}$). The shape of the G mode gives information about the CNTs integrity since an unsymmetrical G mode is characteristic of SWCNTs whereas a symmetric G mode at 1580 cm$^{-1}$ is typical of graphite. Furthermore, graphite also shows a D-band at 1350 cm$^{-1}$. Then, the unsymmetrical G mode and the low $I_D/I_G$ ratio observed in as sintered and deformed 3YTZP/SWCNT composites indicate the absence of severe damages in the CNTs structure after sintering and testing. Radial breathing and G modes in composites exhibits a slightly higher Raman shift than in SWCNT. This effect was observed in zirconia containing different amount of SWCNTs and was attributed to residual stresses in the SWCNTs imposed by the constraining 3YTZP ceramic matrix.[26]

*3.2. High temperature mechanical behaviour*

Creep curves for monolithic 3YTZP and 3YTZP/SWCNT composites are shown in Fig. 3. The steady-state strain rate is clearly reached by the specimens after each stress or temperature change, which allowed us to obtain the stress exponents *n* and the activation energy $Q$, respectively. Comparing both creep curves, it is noticeable that composites are less creep resistant than monolithic 3YTZP. For instance at 1100 ºC, a stress of about 130 MPa should be applied to the later to deform it with a strain rate of ~$10^{-7}$ s$^{-1}$ whereas only 10 MPa is enough for composites to reach this strain rate.

Fig. 3 also shows a lower activation energy Q and higher stress exponent n values in composites than in monolithic 3YTZP. This trend is better observed in Fig. 4. Thus, Fig. 4a shows the log–log plot of the steady-state creep rate versus stress for monolithic 3YTZP and 3YTZP/SWCNTs nanocomposite at 1100, 1150 and 1200 ºC. From the slopes of the linear regressions, average stress exponent values of about 2 and 2.5 are obtained for monolithic 3YTZP and 3YTZP/SWCNTs composites, respectively.

Fig. 4a clearly shows the lower creep resistance exhibited by 3YTZP/SWCNTs nanocomposites compared to monolithic 3YTZP, since the former experience a higher strain rate under the same temperature and stress conditions. It is worth emphasizing that this enhancement in the strain rate cannot be ascribed to the less grain size of nanocomposites with regards to monolithic 3YTZP. In effect, considering in Eq. (1) the grain size data from Table I and taking into account that the grain size exponent $p$ values reported in the literature for this material vary between 1 and 3,[5] then only a correction factor between 1 and 3 is obtained. However, it is experimentally observed a strain rate of at least two orders of magnitude higher for nanocomposites compared to monolithic 3YTZP. The lower creep resistance showed by 3YTZP/SWCNTs nanocomposites compared to monolithic 3YTZP should be due to the weak interfacial bonding between SWCNTs and 3YTZP together with the easy glide within SWCNTs bundles located at the grain boundaries.[21]

Fig. 4b attests that the activation energy Q = 715 ± 60 kJ/mol for 3YTZP specimens experiences a reduction of about 25% by the addition of 2.5 vol.% of SWCNTs, decreasing to Q = 540 ± 40 kJ/mol value . This is evidence that the presence of SWCNTs has a clear influence on the high temperature deformation mechanisms in 3YTZP, which is discussed in the next section.

*3.3. High temperature deformation mechanisms*

As indicated in section 2.1, the residual impurity content of the 3YTZP material studied in this work is higher than 0.1 wt% and then it belongs to the so-called low purity materials.[5,6] Grain boundaries in low-purity 3YTZP materials are coated by an uniform amorphous grain-boundary phase, which becomes liquid like at the high temperatures of the creep tests. The presence of this grain-boundary phase enhances the strain rate since stresses are more easily accommodated than in high purity 3YTZP materials. However, it is worth emphasizing that creep parameters (n and Q) do not depend on the glassy phase.[5] As reported in literature, constant creep parameters (n~2, p~2 and Q~500 kJ/mol) are found over the entire stress range for low purity materials.[6-8] In agreement with these results, for monolithic 3YTZP we found a stress exponent value of about 2, but an activation energy value Q = 715 ± 60 kJ/mol higher than Q~ 500 kJ/mol typically reported in literature for submicron YTZP deformed at temperatures above 1250 ºC.

Gutierrez-Mora *et al.*[30] also reported activation energy values close to 700 kJ/mol in nanocrystalline Y-TZP crept at temperatures between 1150 and 1250 ºC and stresses ranging between 5 and 200 MPa. An interface-controlled mechanism was proposed to account for the experimental results. This mechanism is based on the model of Artz *et al.*[31] and takes into account that grain boundaries are discrete sinks or sources for matter, where point defects are emitted or absorbed at the core of grain-boundary dislocations when they are moving in a non-conservative motion in the grain boundary planes. Thus, the key role of grain-boundary dislocations in this model is to act as a non-perfect sinks or sources of defect points, which enable them to control creep deformation. This model considers that there is a finite number of grain-boundary dislocations whose mobility would be affected by the presence of segregated impurities, and they are evenly spaced in the boundary planes so that they all climb at the same speed. This mechanism could be invoked to justify the high activation energy value obtained from our experiments assuming that 3YTZP is deformed by grain boundary sliding accommodated by grain-boundary dislocation motion, which is controlled by both cation bulk diffusion and local processes taking place at the interface (interface reactions). The annihilation and/or creation of point defects constitutes an additional energetic term that would explain the higher value obtained for the activation energy. However, there are some problems making questionable the application of this model in 3YTZP. In the framework of this model and taking typical values for superplasticity in 3YTZP, an unrealistic number of dislocations in a grain boundary wall can be estimated.[8] Furthermore, an interface-controlled mechanism based on Coble creep would imply changes in the grain morphology in deformed specimens, which are not experimentally observed as shown in table I.

This high activation energy value could come from the yttrium segregation at the grain boundaries, which is well documented in Y-TZP materials.[7-11] A model based on the yttrium segregation effects on high temperature plasticity in YTZP was developed by Gómez-García *et al.*[9] which successfully explain the origin of the threshold stress and its dependence on grain size and temperature. This model considers that yttrium segregation induces a local density of negative charge at the grain boundaries due to the substitution of $Zr^{4+}$ by $Y^{3+}$ cations. This generates a local electric field which is screened by the gradient of oxygen vacancies. According to electrodynamics, the thickness of the layer where the electric field is not zero is in between 1 and 10 nm at room temperature. [9] This local field should influence the $Zr^{4+}$ diffusion, which is the charge carrier controlling diffusion and plastic deformation.[5] Then lattice cationic diffusion as the accommodation mechanism of GBS should contain an additional energetic term due to the effect of yttrium segregation. This term should be added to the bulk cation diffusion energy Q~500 kJ/mol, and then a higher activation energy value is expected. Experimentally we obtain an activation energy Q of about ~715 kJ/mol, i. e. about 50 % higher than Q~500 kJ/mol typically reported in literature. However, the yttrium segregation model predicts a negligible increase in the activation energy for bulk $Zr^{4+}$ diffusion at 1200 ºC in grains with an average size above 0.2 μm.[9]

This disagreement could be explained considering a transition temperature for lattice to grain boundary diffusion (see Fig. 14 in Ref [8]). The effective grain boundary diffusion coefficient depends upon ($\pi\delta/d$), being δ the thickness of the grain boundary diffusion and $d$ the grain size. The accommodation will take place by diffusion of the slowest specie, i.e. $Zr^{4+}$ throughout the easiest path: bulk or grain boundary diffusion. Coming back to Fig. 14 in Ref [8], the intersection point between the bulk and the effective grain boundary diffusion coefficients determines a transition temperature Tc. Above Tc, bulk diffusion coefficient is higher than the grain boundary one, and then GBS is accommodated by bulk diffusion, as it is reported for 3YTZP deformed at temperatures above 1250 ºC.[3-11] However, below Tc grain boundary diffusion coefficient becomes now higher and then it should control creep deformation. We could assume that this is the case at the temperature range of our experiments (1100-1200 ºC). Yttrium segregation at grain boundaries modifies atom chemical surroundings and induces a local electric field which should affect grain boundary diffusion, and then increasing its activation energy. This is coherent to the high activation energy value (Q~715 kJ/mol) obtained from our experiments.

Using the generalised GBS model developed in our group[8], the relative motion of two adjoining grains has two components: parallel (shear motion) defined with a mean flight time $\delta\tau_S$ and perpendicular to their common grain boundary defined with a mean flight time $\delta\tau_D$. Then, $\beta$ can be defined as the ratio of these two characteristic times, whose expression for grain boundary mobility limited by the presence of impurities, grain boundary pinning due to second phase particles or by pores is found to be:[8]

$$\beta = \frac{7\pi \sigma d\, D_{eff}}{24 \gamma\, D_{gb}} \qquad (2)$$

being $\sigma$ the applied stress minus the threshold stress if that exists, $d$ the grain size, $\gamma$ the grain boundary free energy, and $D_{eff}$ is the effective diffusion coefficient along the possible diffusion paths: lattice or grain boundary $D_{gb}$.

An estimation of parameter $\beta$ can be performed relating the effective diffusion coefficient to the grain boundary coefficient by $D_{eff} = (\pi\delta/d)D_{gb}$. So taking typical values $\sigma \sim 50$ MPa, $\gamma \sim 0.15$ J m$^{-2}$ [32], $d \sim 0.3$ μm and $\delta \sim 0.1$ nm then we obtain $\beta \sim 0.09$. The small value obtained for $\beta$ provides a stress exponent value n close to 2 (Eq. 38 in Ref [8]) which is in good agreement to our experimental results.

If grain boundary diffusion controls, then the activation energy will be given by Eq. 42 in Ref [8]:

$$Q = KT + Q_{eff} + \frac{1-\beta}{1+\beta}\left(Q_{eff} - Q_{gb}\right) \qquad (3)$$

where $Q_{eff}$ is the effective activation energy for the combination of simultaneous lattice and grain boundary diffusion, and $Q_{gb}$ is the activation energy for grain boundary diffusion. Since we have estimated $\beta \sim 0.09$ and the term $KT$ is negligible compared to other terms below 2000 K, then Eq. 3 can be written as $Q \approx 2Q_{eff} - Q_{gb}$. $Q_{eff}$ is very close to the activation energy for lattice diffusion $Q_b$ in the limit of large grain size, which is our case since the grain size of 3YTZP specimens is within the submicron scale. Thus, taking $Q_{eff} \approx Q_b \sim 500$ kJ/mol and $Q \sim 715$ kJ/mol from experiments we obtain an activation energy value for grain boundary diffusion $Q_{gb} \sim 300$ kJ/mol. This value seems to be consistent since $Q_{gb}$ is lower than $Q_b$, approximatively $Q_{gb} \approx 0.6 Q_b$, indicating that grain boundary diffusion can be activated at lower temperatures than those for bulk diffusion. Moreover, this value is in very good agreement to those measured by Chaim[33] who found an activation energy value of $280 \pm 10$ kJ mol$^{-1}$ for grain boundary diffusion in 3YTZP at temperatures below 1400 ºC. It is worth noticing that grain growth may be expected in the framework of this model, however the presence of impurities can retard enough grain growth and large deformation can be reached before having grain growth evidences, as it seems to be our case.

In case of 3YTZP/SWCNTs nanocomposites, it is plausible to think that CNTs bundles located around 3YTZP grain boundaries could assist the accommodation and reduction of the local residual stresses. Furthermore, functionalization of CNTs through adsorption of negatively charged carboxyl group, which is used to deagglomerate CNTs via electrostatic repulsion between individual acid-treated CNTs, should affect the yttrium segregation. Thus, negatively charged CNTs located around the 3YTZP grain boundaries should create a negative electric field which could affect yttrium segregation at the grain boundaries by decreasing the segregated yttrium concentration. As a consequence, cation diffusion along grain boundaries is less hindered with regards to that in monolithic 3YTZP. Then, the incorporation of CNTs into a 3YTZP ceramic matrix has a clear influence on the accommodation mechanism. This fact is reflected on the decrease in the activation energy value from Q = 715 ± 60 kJ/mol for monolithic 3YTZP to Q = 540 ± 40 kJ/mol for 3YTZP/SWCNTs nanocomposites, which is close to the typical activation energy value Q~500 kJ/mol reported in literature for low purity Y-TZP.[5-6] Studies on the yttrium distribution in 3YTZP grain to investigate how CNTs could affect yttrium segregation at the grain boundary are required to support this explanation.

## 4. Conclusions

High temperature mechanical behaviour have been investigated in monolithic 3YTZP and 3YTZP/SWCNTs nanocomposites, being the latter less creep resistant. Creep parameters and microstructural observations of specimens prior and after creep tests point out grain boundary sliding (GBS) as the high temperature deformation mechanism in both materials. The activation energy values obtained under our experimental conditions suggest that GBS is accommodated by grain boundary diffusion in monolithic 3YTZP specimens, where yttrium segregation at grain boundaries induces a local electric field which increases the activation energy for GB diffusion. In 3YTZP/SWCNTs nanocomposites, SWCNTs bundles around 3YTZP grains should influence yttrium segregation, being grain boundary diffusion less hindered compared to monolithic 3YTZP and consequently obtaining a lower activation energy.

**Acknowledgements**

This work was financially supported by the European Regional Development Fund and the Spanish "Ministerio de Economía y Competitividad" through the projects MAT2009-11078, MAT2012-34217 and the project from the Andalucia Government P12-FQM-1079. M. C-R thanks the JAE-doc contract awarded by the Spanish CSIC, co-financed by the European Social Fund.

**FIGURE CAPTIONS**

Fig. 1. HRSEM micrographs of a) polished and thermally etched surface of monolithic 3YTZP, and fracture surfaces of b) 3YTZP/SWCNTs composite and c) 3YTZP/SWCNTs composite after creep test.

Fig. 2. Raman spectra of monolithic 3YTZP, SWCNTs, 3YTZP/SWCNTs nanocomposites before and after creep tests.

Fig. 3. Creep curves, showing the stress and temperature changes to obtain the stress exponent n and the activation energy Q, for a) monolithic 3YTZP and b) 3YTZP/SWCNTs nanocomposites.

Fig. 4. a) Strain rate-stress logarithm plot for monolithic 3YTZP and 3YTZP/SWCNTs nanocomposites. The average stress exponent n for each temperature is shown. b) Strain rate in logarithmic scale versus 1/T plot indicating the activation energy obtained for monolithic 3YTZP and 3YTZP/SWCNTs nanocomposites.

**TABLES**

Table I. Relative density $\rho_r$ and morphological parameters (grain size $d$ and shape factor $F$) for both 3YTZP and 3YTZP-SWCNTs, as-sintered and after creep tests.

| Material | $\rho_r \pm 0.5$ (%) | 3YTZP grains | |
|---|---|---|---|
| | | $d$ (μm) | $F$ |
| 3YTZP | 99.5 | 0.27 ± 0.10 | 0.72 ± 0.07 |
| 3YTZP deformed | | 0.25 ± 0.12 | 0.74 ± 0.07 |
| 3YTZP-SWCNTs | 99.7 | 0.20 ± 0.09 | 0.71 ± 0.09 |
| 3YTZP-SWCNTs deformed | | 0.20 ± 0.08 | 0.69 ± 0.08 |

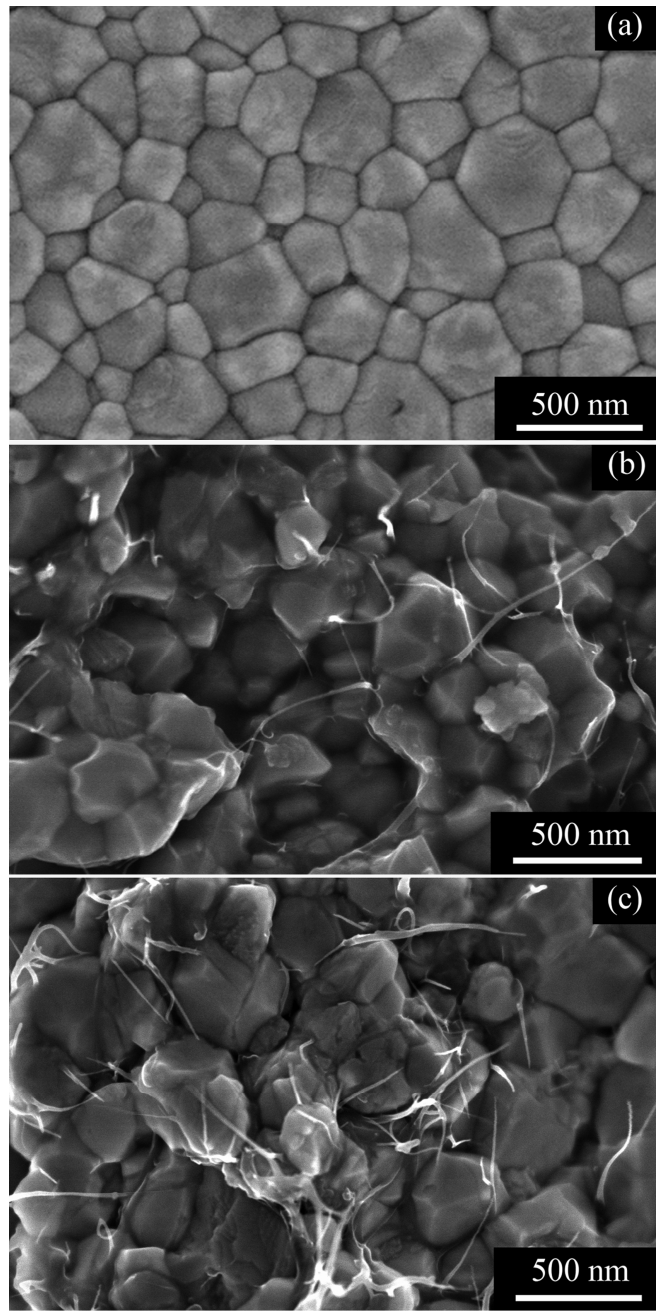

Fig. 1.

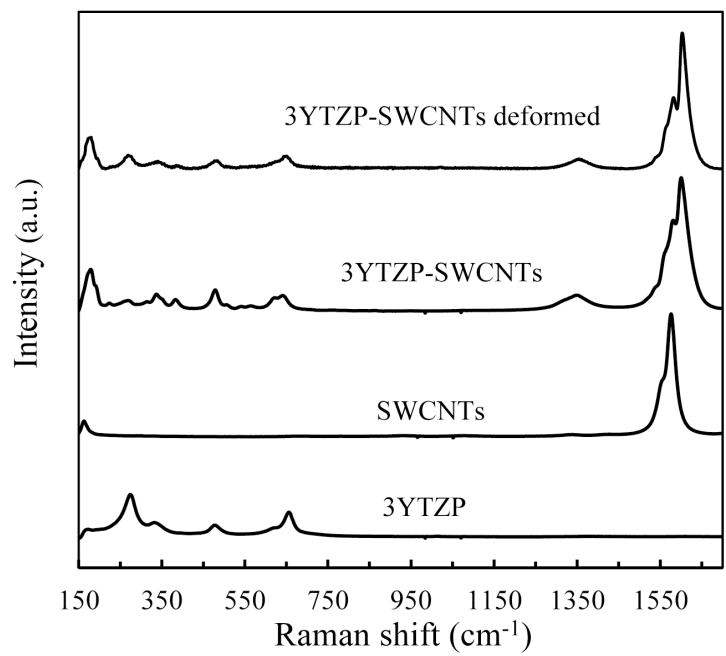

Fig. 2.

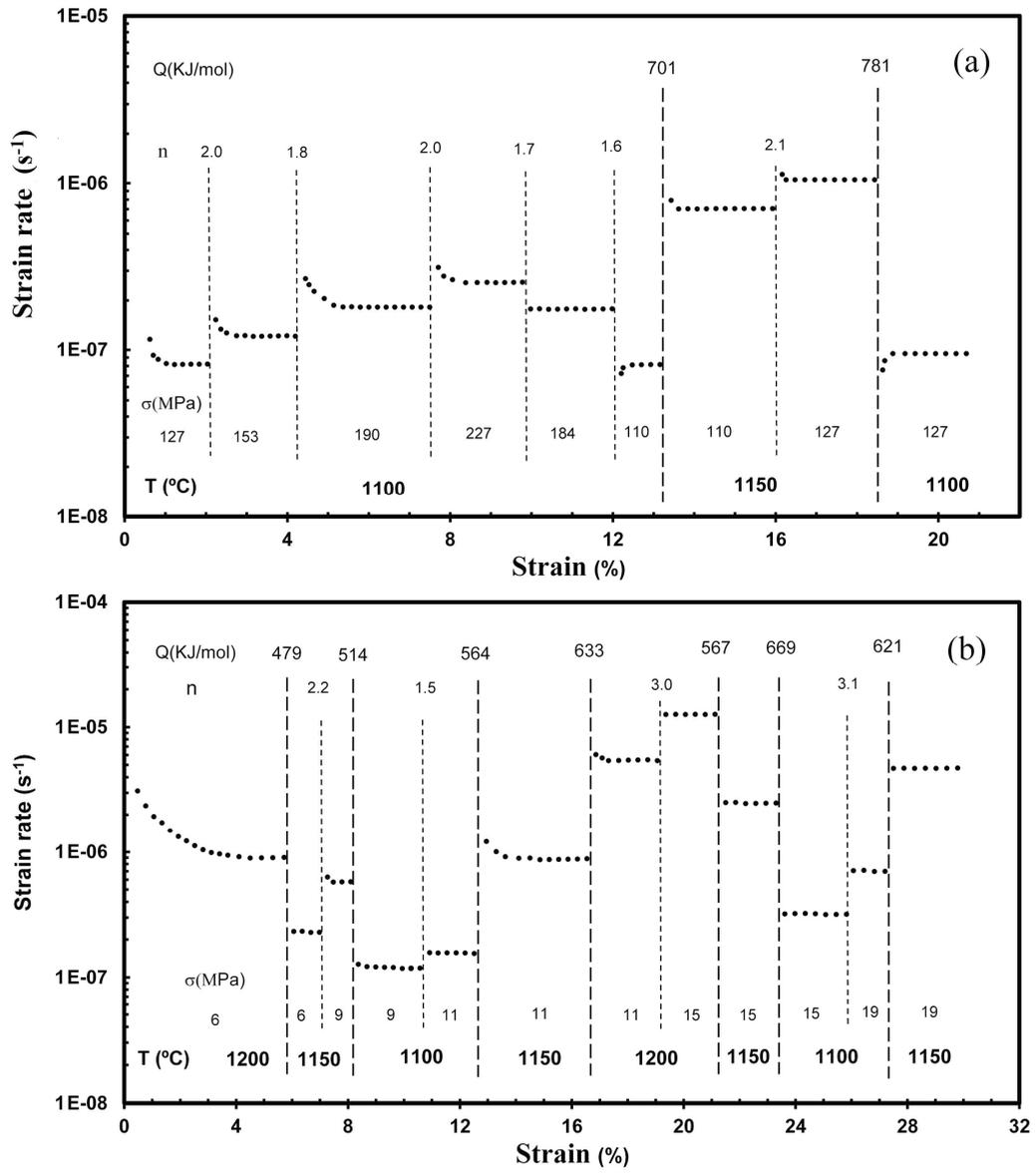

Fig. 3.

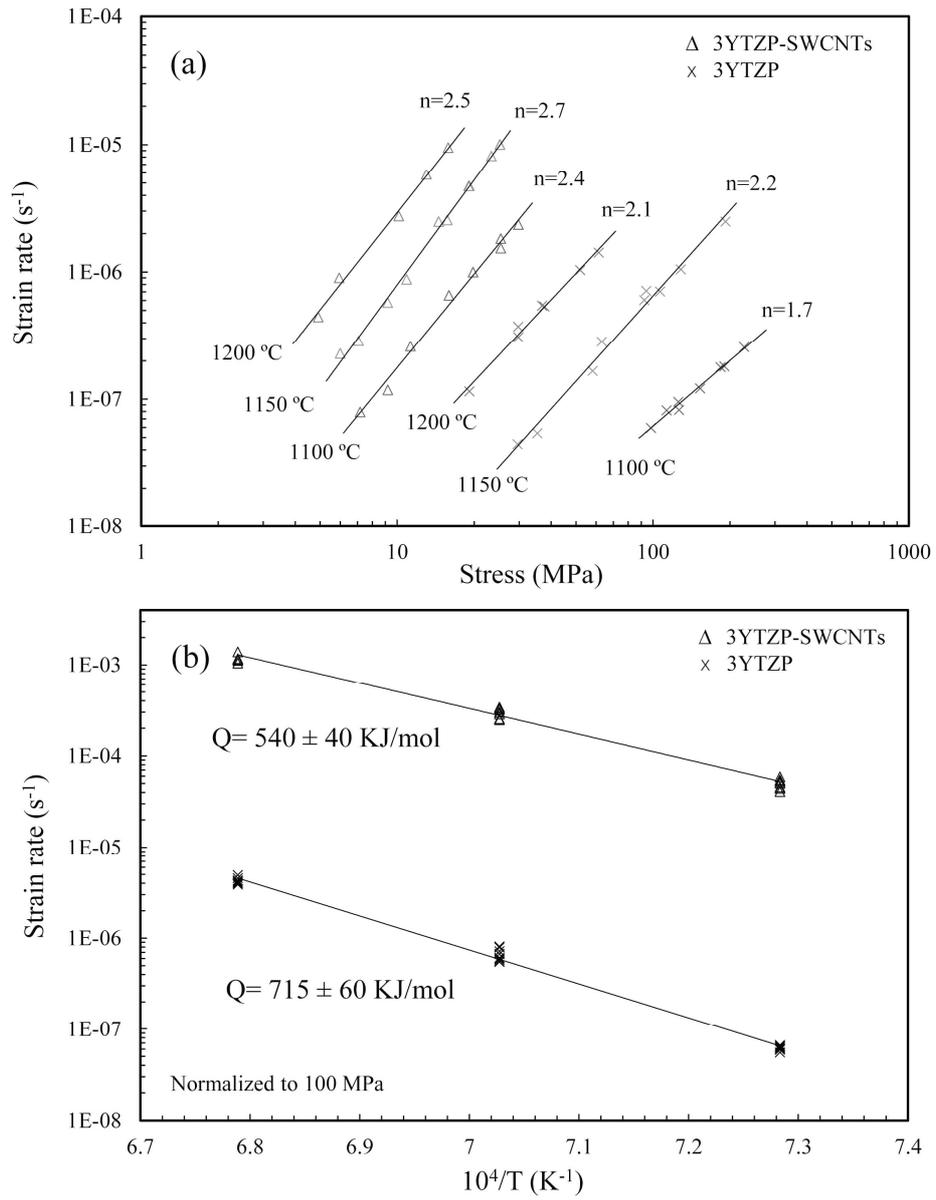

Fig.4.